\documentclass[onecolumn]{rps-esrel2022b}

\def\papername{\jobname}

\usepackage{natbib,tikz,import,float}

\begin{document}

\markboth{T.M. Julitz, A. Tordeux and M. Löwer}{Reliability of fault-tolerant system architectures for automated driving systems}


\title{\bf Reliability of fault-tolerant system architectures for automated driving systems}

\author{Tim M. Julitz}

\address{Chair for Product Safety and Quality Engineering, University of Wuppertal, Germany. \email{julitz@uni-wuppertal.de}}

\author{Antoine Tordeux}

\address{Chair for Reliability and Traffic Safety, University of Wuppertal, Germany. \email{tordeux@uni-wuppertal.de}}

\author{Manuel Löwer}

\address{Chair for Product Safety and Quality Engineering, University of Wuppertal, Germany. \email{loewer@uni-wuppertal.de}}

\begin{abstract} 
Automated driving functions at high levels of autonomy operate without driver supervision.
The system itself must provide suitable responses in case of hardware element failures.
This requires fault-tolerant approaches using domain ECUs and multicore processors operating in lockstep mode. 
The selection of a suitable architecture for fault-tolerant vehicle systems is currently challenging.
Lockstep CPUs enable the implementation of majority redundancy or M-out-of-N ($M$oo$N$) architectures. 
In addition to structural redundancy, diversity redundancy in the ECU architecture is also relevant to fault tolerance. 
Two fault-tolerant ECU architecture groups exist: architectures with one ECU (system on a chip) and architectures consisting of multiple communicating ECUs. 
The single-ECU systems achieve higher reliability, whereas the multi-ECU systems are more robust against dependent failures, such as common-cause or cascading failures, due to their increased potential for diversity redundancy. 
Yet, it remains not fully understood how different types of architectures influence the system reliability. 
The work aims to design architectures with respect to CPU and sensor number, $M$oo$N$ expression, and hardware element reliability. 
The results enable a direct comparison of different architecture types. 
We calculate their reliability and quantify the effort to achieve high safety requirements.
Markov processes allow comparing sensor and CPU architectures by varying the number of components and failure rates. 
The objective is to evaluate systems' survival probability and fault tolerance and design suitable sensor-CPU architectures.
The results show that the system architecture strongly influences the reliability.
However, a suitable system architecture must have a trade-off between reliability and self-diagnostics that parallel systems without majority redundancies do not provide. 
\end{abstract}




\section{Introduction}
The draft law amending the road traffic code and the compulsory vehicle insurance in Germany takes the development of automated vehicle systems to the next level \citep{BMVI2021}.
The law creates the conditions for the use of highly automated vehicles (SAE level 4, \citet{J3016_202104}) in public road traffic. 
Already in 2017, the eighth amendment to the road traffic code came into force, enabling the operation of level 3 vehicles.
The beginnings of research on vehicle automation go back to the PROMETHEUS project, which started in 1986 and laid the foundations of today's commercial driver assistance systems up to level 2 \citep{williams1988}. Some examples are the distance cruise control Distronic Plus and the emergency brake assistant Pre-Safe from Daimler. The next milestones in the development of autonomous vehicles were achieved with the launch of the DARPA Challenges in 2004. In 2005, driverless vehicles made their way over 212 km through the Mojave Desert, focusing on autonomous navigation \citep{crane2007}. In the follow-up project in 2007, urban traffic was simulated on an abandoned Air Force base \citep{buehler2009}. 

A large number of software architectures emerged from the DARPA Challenges, which have one essential thing in common: The architectures are divided into modules that fulfil different functions. 
The modules essentially consist of localisation, perception and vehicle control \citep{reke2020}. 
The 2005 winning vehicle team identified some significant problems. 
The developed vehicle was able to successfully drive in a static environment, but navigation through road traffic is not possible due to the insufficient reliability of the system \citep{thrun2006stanley}. 
The hardware architecture of the vehicle consisted of six computers that performed various functions. 
Watchdogs monitored the states of software and hardware to restart the system in case of failure. 

During the 2007 DARPA Challenge, significant successes were achieved in the monitored urban environment. 
The first-place vehicle system relied on different modes of operation: a normal state and a recovery state \citep{urmson2008autonomous}. 
The recovery state is triggered when objects block the planned path, objects are detected too late, or actions are kinematic infeasible. 
Four algorithms are used to return to normal operation, which essentially consist of replanning paths and increasing the safety margin \citep{urmson2008autonomous}. 
The software-based solutions allow increasing robustness. 
The hardware-based measures consist of a dual-core CPU and the combination of various sensors. 
However the reliability and robustness of the vehicle was not sufficient to drive in real road traffic, which is considerably more complex than the monitored environment. 
Although the vehicle was able to recover from many fault conditions, the time required to do so was considerable, up to ten minutes, which is not reasonable in the real environment. 
Therefore, the developed system cannot be classified with SAE level 4.

In the last ten years, many projects from industrial and academic organisms have further advanced the state of the art. 
On the way to SAE Level 4 driverless operation \cite{J3016_202104}, repeated accidents of autonomous vehicles show that further research is needed to increase the safety, reliability and robustness of the automated driving systems \cite{daily2017self}. 
The elimination of the human fallback level requires fault-tolerant approaches to system modelling that cannot be implemented by simple redundancy. 
The principles of fault tolerance are based on self-diagnosis, reliability and availability evaluation, reconstruction and error recovery. 
The reliability is usually increased by structural redundancy. 
Availability indicates whether a system is functioning at a certain point in time and can be influenced by diversity redundancy, operation independence, or asymmetric system architectures. 
These principles are already applied in traditional safety-critical systems, which can be found in aviation, rail transport, space travel, military or nuclear power plants, among others. 
They are also becoming increasingly important in the automotive literature, see, e.g., \cite{baleani2003fault,kohn2015fail,ishigooka2018cost,schmid2019safety,lin2018architectural,sari2020fail}.
The development of a fault-tolerant system architecture remains nowadays one of the most important challenges for the market introduction of autonomous vehicles \cite{daily2017self}. 

In this contribution, we aim to analyse and compare different types of redundant and fault-tolerant architectures including parallel and $M$oo$N$ systems using Markovian processes.
The contribution is organised as following.
We present the types of considered architecture models in Sec.~\ref{sec:2}. 
The modelling and analysis of the models using Markovian processes are detailed in Sec.\ref{sec:3}. 
Numerical results for different types of architectures are presented in Sec.~\ref{sec:4} and discussed in Sec.~\ref{sec:5}.

\section{Architecture models \label{sec:2}}

\begin{figure}[!ht]
\centering
\vspace{1em}
\def\svgwidth{0.5\textwidth}\small
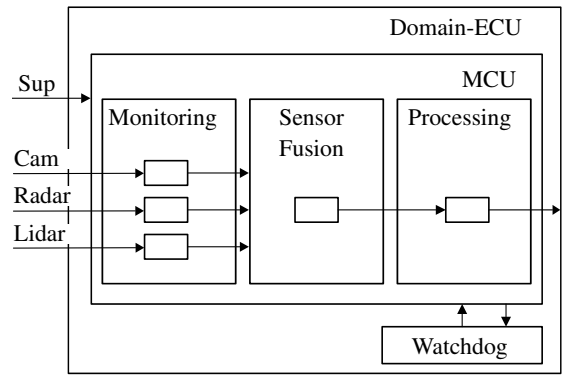
\vspace{0em}
\caption{Structure of a MCU in 1oo1 architecture based on \cite{sari2020fail}}
\label{fig:MCU}
\end{figure}

Many different redundant system architectures exist in the literature. 
In the databases of Springer, the IEEE Xplore Digital Library, the Wiley Online Library and the ACM Digital Library, the keywords "fail operational", "autonomous driving architecture", "fallback strategy", "$M$oo$N$ redundancy", "system on a chip", "reliability" and "functional safety" are currently frequently used. 
Indeed, $M$oo$N$ redundancy configurations allow flexible operation of systems, suitable with recovery and self-diagnose processes. 
These characteristics are well adapted to the automation of complex systems, and especially automation of the driving. 
In \cite{sari2020fail}, the hardware components of a fault-tolerant system architecture is defined as a combination of microcontroller units (MCUs) supplied by three independent sensors (camera, radar, lidar), both configurate in a $M$oo$N$ redundancy. The structure of a single MCU is shown in fig ~\ref{fig:MCU} which will be used as a basis for the following.

The MCUs are powered by independent power supplies and are controlled by independent watchdogs. They consists of three blocks. A monitoring unit, a sensor fusion and a processing block, which are supplied with information from the sensors. This is the 1oo1 MCU architecture which is embedded in a Domain-Electronic-Control-Unit (Domain-ECU) and provides the basis for various $M$oo$N$ redundancies. In addition, the following distinguishes between a sensor architecture and a MCU architecture, which is noted as $S$oo$N_S$/$M$oo$N_M$. The sensor architecture specifies how many of the available sensors must be operational to ensure the functionality of the system.

The selection of the sensor architecture represents a central research topic in the development of autonomous vehicles \cite{daily2017self}. 
Two types of architectures are identified: 
\begin{enumerate}
    \item System-on-a-Chip architectures (hereafter referred to as 1-ECU system) and 
    \item Dual-Fail-Safe architectures (hereafter referred to as 2oo2DFS system).
\end{enumerate} 
The basic 1-ECU architecture includes a 2oo3 redundancy that requires the operation of four MCUs on one ECU. 
There are three independent processing strings for the sensor information (MCU 1 to 3), each supplied by an independent power supply (Sup 1 to 3). 
The information comes from the same sensors (camera, radar, lidar). 
These are not additionally redundant. 
The redundancy lies in the diversification of different sensors. 
Furthermore, a majority decision takes place in MCU 4. 
For this purpose, the outputs of MCUs 1 to 3 are redundantly fed into processor cores 1 to 3 of MCU 4, which execute the 2oo3 majority comparison \cite{sari2020fail}. 
Subsequently, the 2oo3 process is repeated a second time by core 4 before the result is passed on as output. The second type of architecture consists of the 2oo2DFS architecture shown in \citet{sari2020fail}.
The 2oo2DFS architecture consists of two independent fail-safe subsystems that are outsourced to different ECUs and communicate with each other in isolation. In the event of a fault, one ECU takes over the function of the other. 
In this operating state, the system no longer runs fail-operationally, but only fail-safe as long as both ECUs are not restored. 
The two fail-safe subsystems are operated via two independent MCUs, which process the sensor information and are in 2oo2 majority comparison with each other. 
A third MCU implements this comparison.

\section{Markovian analyse \label{sec:3}}

\begin{figure}[!ht]
\centering
\bigskip
\includegraphics[width=.5\textwidth]{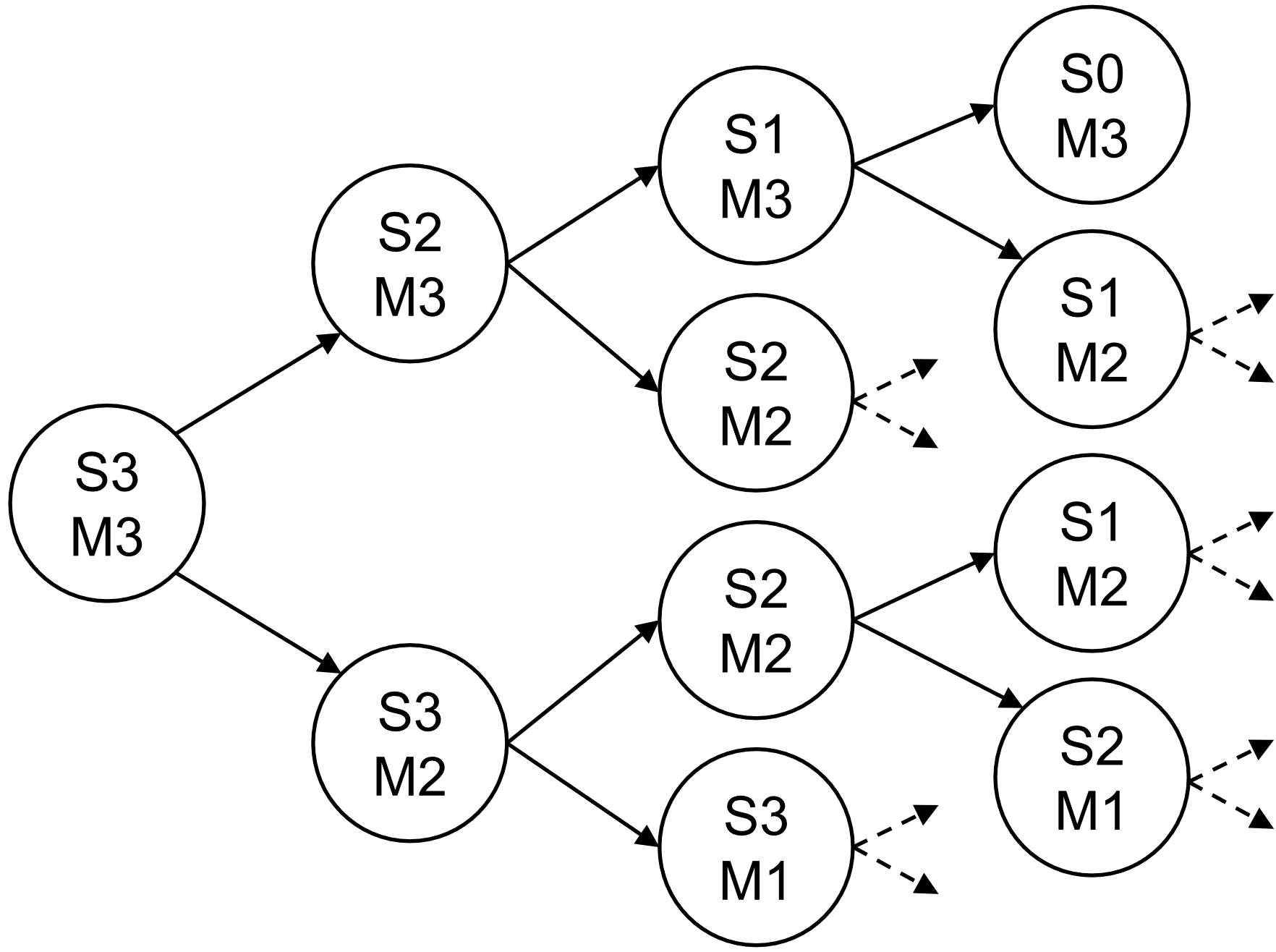}
\caption{Phase diagram of a Markov chain for a system with three MCUs and three sensors}
\label{fig_markov}
\end{figure}

The selection of a suitable architecture for fault-tolerant functions is a current challenging research topic. 
Various approaches exist. 
The differences lie mainly in the number of MCUs, the number of sensors and the degree of majority redundancy. 
For example, the 1-ECU architecture can also be operated with 2 sensors and 4 MCUs in the 2oo4 majority comparison. 
Using Markovian processes, we can determine the reliability of the systems, compare different sensor and MCU architectures, and determine which is the most suitable. 
The phase diagram of a Markov chain for a system with three MCUs and three sensors is outlined in Fig.~\ref{fig_markov}.

We consider in the following a system with $N_M$ identical MCUs and $N_S$ identical sensors. 
The MCU and sensor times to failure are respectively identically exponentially distributed with $\lambda_M>0$ the failure rate of a MCU and $\lambda_S>0$ the failure rate of a sensor.
We denote $P_{m,s}(t)$ the probability that $m=0,\ldots N_M$ MCU(s) and $s=0,\ldots,N_S$ sensor(s) are operating at time $t$. 
We have $\sum_{m,s}P_{m,s}(t)=1$ for all $t$. 
The definition of the failure of the entire system depends on the majority redundancy used. A 2oo3 sensor and 3oo4 MCU architecture is operational when at least two sensors and three MCUs are functional.
We consider in the following that the MCUs and sensors in, respectively, $M$oo$N_M$ and $S$oo$N_S$ configurations. 
This means that the system fails as soon as the number of operating MCU(s) $m$ is strictly less than $M$ or the number of operating sensor(s) $s$ is strictly less than $S$. 
Therefore the reliability of the system is
\begin{equation}
    R(t)=\sum_{m=M}^{N_M}\sum_{s=S}^{N_S} P_{m,s}(t).
    \label{R}
\end{equation}

There exist $(N_M+1)(N_S+1)$ possibles states. 
However, only few transitions from a state to another are possible. 
Indeed, the time being continuous, only one sensor or MCU can fail at a given instant. 
Furthermore a failed component can not be restored.
Therefore only transitions $(m,s)\mapsto(m-1,s)$, $m>0$, and $(m,s)\mapsto(m,s-1)$, $s>0$, are possible.
Consequently, the transition matrix of the Markov process associated to the $M$oo$N$ systems is triangular, and mostly sparse. 
For instance, the transition matrix $A$ associated to an architecture consisting of three sensors and three MCUs as described in Fig.~\ref{fig_markov} is given by
\begin{equation}
\label{A}
A=\text{\footnotesize
\def\ee{\hspace{-2mm}}
$\left[\begin{array}{cccccccccccc}
-2\lambda_S-3\lambda_M\ee&3\lambda_M&0&0&2\lambda_S&0&0&0&0&0&0&0\\
0&\ee-2\lambda_S-2\lambda_M\ee&2\lambda_M&0&0&2\lambda_S&0&0&0&0&0&0\\
0&0&\ee-2\lambda_S-\lambda_M&\lambda_M&0&0&2\lambda_S&0&0&0&0&0\\
0&0&0&-2\lambda_S&0&0&0&2\lambda_S&0&0&0&0\\
0&0&0&0&-\lambda_S-3\lambda_M\ee&3\lambda_M&0&0&\lambda_S&0&0&0\\
0&0&0&0&0&\ee-\lambda_S-2\lambda_M\ee&2\lambda_M&0&0&\lambda_S&0&0\\
0&0&0&0&0&0&\ee-\lambda_S-\lambda_M&\lambda_M&0&0&\lambda_S&0\\
0&0&0&0&0&0&0&-\lambda_S&0&0&0&\lambda_S\\
0&0&0&0&0&0&0&0&-3\lambda_M&3\lambda_M&0&0\\
0&0&0&0&0&0&0&0&0&-2\lambda_M&2\lambda_M&0\\
0&0&0&0&0&0&0&0&0&0&-\lambda_M&\lambda_M\\
0&0&0&0&0&0&0&0&0&0&0&0
\end{array}\right]$}
\end{equation}
We denote the vector of probabilities of the different possible states
\begin{equation}
\begin{array}{l}
    P=(P_{N_M,N_S},P_{N_M-1,N_S},\ldots,P_{0,N_S},\\
    \hspace{.65cm} P_{N_M,N_S-1},P_{N_M-1,N_S-1},\ldots,P_{0,0})^\text T.
    \end{array}
\end{equation}
The state probabilities $P$ of the system and, using Eq.~(\ref{R}), its reliability, 
are solution of the Kolmogorov backward linear differential equation
\begin{equation}
    \dot P(t)=A^\text T P(t),\qquad P(0)=P_0.
\end{equation}
Here $P_0$ is the initial state of the system.
We obtain directly the solution
\begin{equation}
    P(t)=e^{A^\text Tt}P_0.
\end{equation}
The calculation of the matrix exponential can be performed iteratively since the matrix $A$ is triangular. More generally, the matrix exponential can be obtained using linear algebra and matrix diagonalisation, Laplace transform, or numerically on a computing software.

By varying the number of MCUs, sensors and their failure rates $\lambda_M$ and $\lambda_S$ parameters, the system configuration can be adapted with regard to reliability-related criteria. 
The aim is to identify the configuration of the system architecture with the greatest probability of survival with sufficient self-diagnosis function. 
The self-diagnosis consists of selecting a suitable majority redundancy. 
Systems with a high proportion of series ($N$oo$N$) and pure parallel systems (1oo$N$) do not have any self-diagnosis functions without the use of additional hardware designed for this purpose and are therefore not considered. 
For cost reasons, the number of sensors is generally limited to a maximum of three (cf. camera, radar, lidar). Additional sensors such as ultrasound do not serve to ensure the safety objective of the driving task. The number of MCUs is limited to a maximum of four. This makes it possible to realise up to 3oo4 architectures known from space travel or nuclear power plants. 

\section{Numerical results \label{sec:4}}

\begin{figure*}[!ht]
\centering
\input{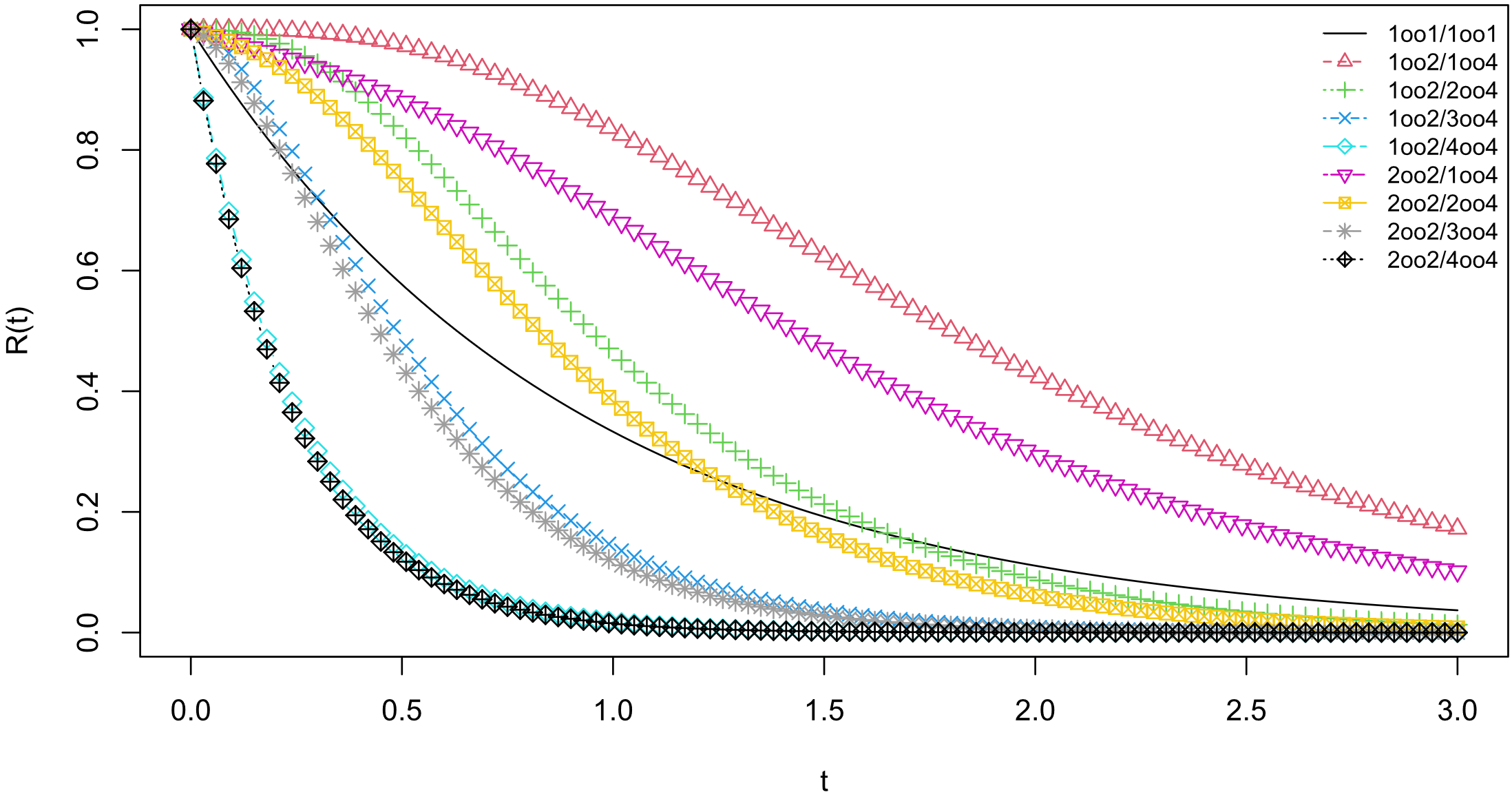}\vspace{-2mm}
\caption{Survival probability $R(t)$ of different majority redundant $S$oo$N_S$/$M$oo$N_M$ sensor/MCU architectures with $N_S=3$ sensors and $N_M=3$ MCUs. $\lambda_S=0.1~\cdot~10^{-4}~$h$^{-1}$ and $\lambda_M=1~\cdot~10^{-4}~$h$^{-1}$.}
\label{fig:R1}\bigskip
\end{figure*}

\begin{figure*}[!ht]
\centering
\input{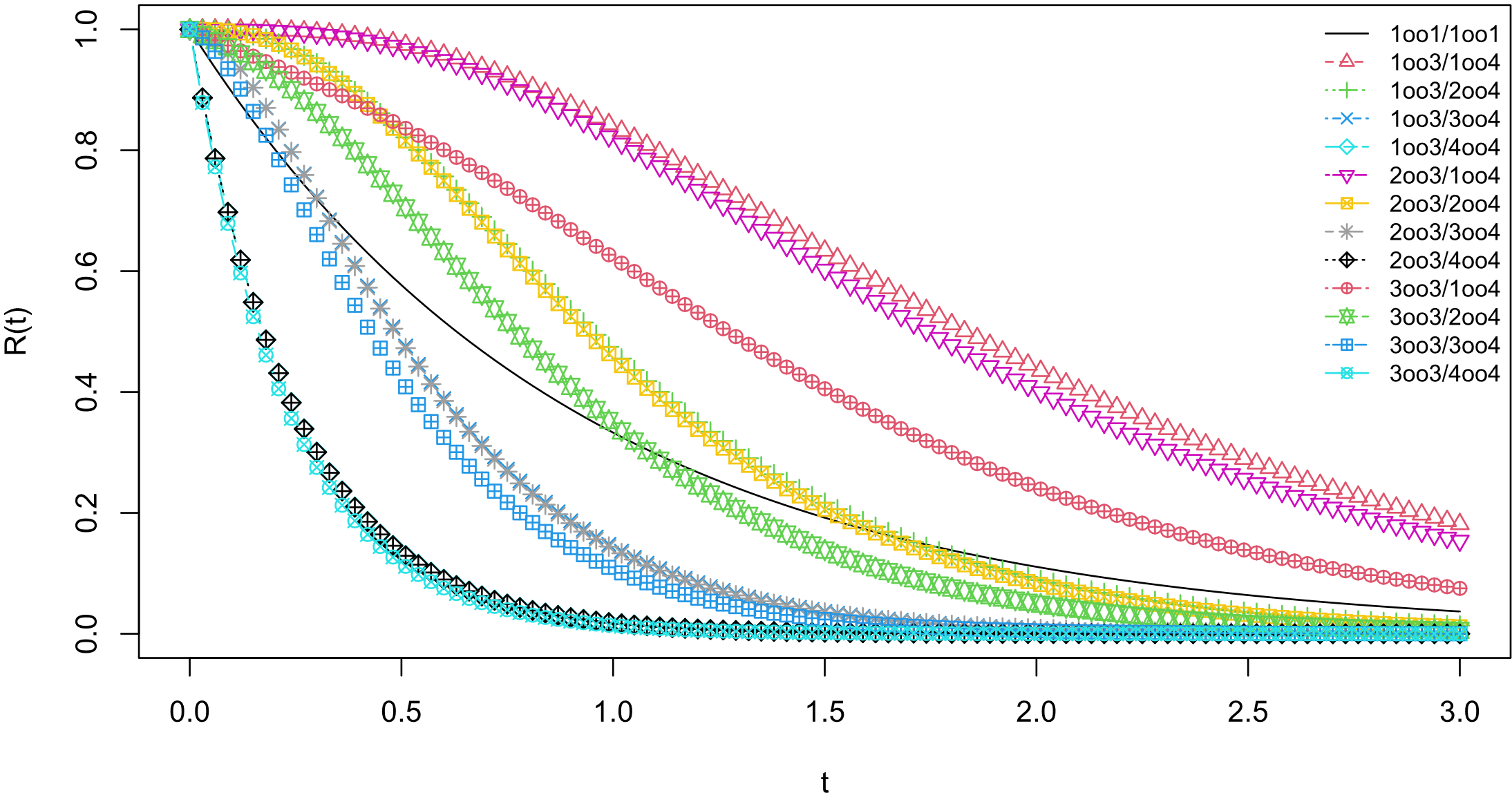}\vspace{-2mm}
\caption{Survival probability $R(t)$ of different majority redundant $S$oo$N_S$/$M$oo$N_M$ sensor/MCU architectures with $N_S=3$ sensors and $N_M=4$ MCUs. $\lambda_S=0.1~\cdot~10^{-4}~$h$^{-1}$ and $\lambda_M=1~\cdot~10^{-4}~$h$^{-1}$.}
\label{fig:R2}\bigskip\medskip
\end{figure*}

Within the Markovian framework, any majority sensors/MCUs configuration may be considered.
In the following, we analyse numerically systems consisting of up to three sensors and four MCUs (S3M4). 
We set $\lambda_S=0.1~\cdot~10^{-4}~$h$^{-1}$ and $\lambda_M=1~\cdot~10^{-4}~$h$^{-1}$ and assume that all components alre initially operating. 
Figure \ref{fig:R1} shows the survival probability $R(t)$ (reliability) of the combination of different majority redundant systems of sensors and MCUs for an architecture consisting of three sensors and three MCUs.
The leading designation $M$oo$N_S$ describes the majority redundancy of the sensors and the trailing designation $M$oo$N_M$ the majority redundancy of the MCUs. 
A simple series system without redundancy 1oo1/1oo1 is listed as a reference (black solid curve). 
Interestingly, the failure rate of the sensors must be selected lower than the failure rate of the MCUs in order to achieve higher reliability values. 
The 3oo3/3oo3 and 2oo3/3oo3 systems have the lowest reliability. 
These are systems with a large proportion of series. 
The two curves with the highest reliability are assigned to the 1oo3/1oo3 and the 2oo3/1oo3 system. 
However, these are also not suitable for fault-tolerant operation, as they are systems with an exclusive or large parallel share (1oo$N$).
Parallel systems have no self-diagnosis functions. 
They are not able to validate their input independently through majority redundancy. 
The focus is therefore on the middle curves in the area of the reference line. 
Of these, the grey curve (3oo3/1oo3) has the greatest probability of survival. 
However, due to the combination of a pure series and a pure parallel system, it must also be sorted out. 
Taking into account the self-diagnosis property, the 2oo3/2oo3 architecture of the S3M3 systems can be said to be the most suitable for fault-tolerant operation. 
However, it must be noted that the curve falls short of the 1oo1/1oo1 reference in the average course of the service life.

Fig.~\ref{fig:R2} shows the same investigation with four MCUs.
Once again, the curves in the upper and lower reliability range correspond to systems with large series and large parallel parts with low self-diagnosis capacity, which are to be sorted out. 
The curves with suitable majority redundancy are again in the range of the reference system 1oo1/1oo1. 
This includes the 3oo3/2oo4 and the 1oo3/2oo4 systems, which are also parallel or series systems. 
It remains mostly the 2oo3/2oo4 system. 
Compared to the 2oo3/2oo3 system of the S3M3 architecture, the 2oo3/2oo4 can be said to be even more suitable, since the crossing with the reference system 1oo1/1oo1 occurs much later.

\section{Discussion \label{sec:5}}

The introduction of automated or autonomous vehicle systems of automation SAE levels four and five requires the development of fault-tolerant system architectures allowing to dispense with the driver's fallback level. 
In this contribution, different fault-tolerant approaches by majority systems were analysed and system properties were derived that can be used for further developments. 
Research so far has focused on separate aspects of fault tolerance based on self-diagnosis, reliability, availability, reconstruction and fault recovery. 
The present analysis combines the areas of self-diagnosis and reliability.

The highest reliability is achieved by pure parallel systems (1oo$N$). 
However, they are not suitable for fault-tolerant applications due to their lack of self-diagnostic capability. 
$M$oo$N$ majority redundancies with M\,$> 1$ validate their input by a majority comparison, 
Whereby in the event of a fault, subsystems can be switched off and different operating modes can be used depending on the system state. The development of diagnostic and monitoring facilities for fault-tolerant operation of 1oo$N$ systems would advance the development of fault-tolerant system architecture in terms of reducing complexity and increasing reliability.

2oo4 architectures are considerably more reliable than 2oo3 architectures. 2oo4 systems are mainly established in the military sector, in space travel and in nuclear power plants. For the automotive sector, however, mainly only 2oo2 and 2oo3 architectures are discussed. It is recommended that future research should increasingly consider 2oo4 systems. An important question to be addressed is whether the reliability gain can be justified in comparison to the increased costs due to the increased number of components and in comparison to the greater hardware overhead.
If a system based on a single domain ECU is used, then a 2oo3 sensor/ 2oo4 MCU architecture or a 2oo2 sensor/ 2oo4 MCU architecture provides the best performance in terms of reliability, taking into account its self-diagnostic capability. 
Systems with a large parallel component deliver the highest reliability and systems with a high serial component deliver the lowest reliability. 
The reliability changes only marginally between sensor architectures 2oo2 and 2oo3. From a reliability point of view, it therefore plays a subordinate role whether two or three sensors are used.

The components of the system architectures can be divided into three groups for their design to achieve reliability target values (e.g., ISO 26262 \cite{ISO26262}), which is essentially based on the order of the minimum sections associated with the components. In the application case in question, a factor of approx. 10 between the failure rate of components of the sensor and the MCU architecture has proven itself. This finding is transferable to further system designs. It should be noted that the minimum cuts cannot be blindly used as a basis, since various effects can lead to a component being more critical with regard to a system failure than the other representatives of its cut-set order. For this purpose, the use of importance analyses can be recommended.

For further work, it makes sense to also deal with the other aspects of availability, reconstruction and fault recovery. The DFS architecture presented is ideally suited for this purpose. To reduce complexity, the two ECUs of the DFS system have been actively redundant. However, real fault tolerance can only be achieved through passive redundancy or stand-by redundancy. This way, the aspect of reconstruction is taken into account. In the event of a fault, ECU 1 switches off and ECU 2 takes over operation. Meanwhile, error correction algorithms (e.g. error correcting codes) restore the failed components of ECU 1, thus taking the aspect of error correction into account. The time to repair a component can be implemented as a repair rate, which allows availability considerations. Through this approach, all aspects of fault tolerance can be covered.




\begin{thebibliography}{}

\bibitem[\protect\citeauthoryear{Baleani, Ferrari, Mangeruca,
  Sangiovanni-Vincentelli, Peri, and Pezzini}{Baleani
  et~al.}{2003}]{baleani2003fault}
Baleani, M., A.~Ferrari, L.~Mangeruca, A.~Sangiovanni-Vincentelli, M.~Peri, and
  S.~Pezzini (2003).
\newblock Fault-tolerant platforms for automotive safety-critical applications.
\newblock In {\em Proceedings of the 2003 international conference on
  Compilers, architecture and synthesis for embedded systems}, pp.\  170--177.

\bibitem[\protect\citeauthoryear{Buehler, Iagnemma, and Singh}{Buehler
  et~al.}{2009}]{buehler2009}
Buehler, M., K.~Iagnemma, and S.~Singh (2009).
\newblock {\em {The DARPA urban challenge: autonomous vehicles in city
  traffic}}, Volume~56.
\newblock springer.

\bibitem[\protect\citeauthoryear{Crane}{Crane}{2007}]{crane2007}
Crane, C.~D. (2007).
\newblock {The 2005 DARPA Grand challenge}.
\newblock In {\em 2007 International Symposium on Computational Intelligence in
  Robotics and Automation}, pp.\  nil4--nil4. IEEE.

\bibitem[\protect\citeauthoryear{Daily, Medasani, Behringer, and Trivedi}{Daily
  et~al.}{2017}]{daily2017self}
Daily, M., S.~Medasani, R.~Behringer, and M.~Trivedi (2017).
\newblock Self-driving cars.
\newblock {\em Computer\/}~{\em 50\/}(12), 18--23.

\bibitem[\protect\citeauthoryear{Ishigooka, Honda, and Takada}{Ishigooka
  et~al.}{2018}]{ishigooka2018cost}
Ishigooka, T., S.~Honda, and H.~Takada (2018).
\newblock Cost-effective redundancy approach for fail-operational autonomous
  driving system.
\newblock In {\em 2018 IEEE 21st International Symposium on Real-Time
  Distributed Computing (ISORC)}, pp.\  107--115. IEEE.

\bibitem[\protect\citeauthoryear{??}{ISO 26262:2018}{2018}]{ISO26262}
ISO 26262:2018 (2018).
\newblock {Road vehicles — Functional safety}.
\newblock Standard, International Organization for Standardization, Geneva, CH.

\bibitem[\protect\citeauthoryear{Kohn, K{\"a}{\ss}meyer, Schneider, Roger,
  Stellwag, and Herkersdorf}{Kohn et~al.}{2015}]{kohn2015fail}
Kohn, A., M.~K{\"a}{\ss}meyer, R.~Schneider, A.~Roger, C.~Stellwag, and
  A.~Herkersdorf (2015).
\newblock Fail-operational in safety-related automotive multi-core systems.
\newblock In {\em 10th IEEE International Symposium on Industrial Embedded
  Systems (SIES)}, pp.\  1--4. IEEE.

\bibitem[\protect\citeauthoryear{Lin, Zhang, Hsu, Skach, Haque, Tang, and
  Mars}{Lin et~al.}{2018}]{lin2018architectural}
Lin, S.-C., Y.~Zhang, C.-H. Hsu, M.~Skach, M.~E. Haque, L.~Tang, and J.~Mars
  (2018).
\newblock The architectural implications of autonomous driving: Constraints and
  acceleration.
\newblock In {\em Proceedings of the Twenty-Third International Conference on
  Architectural Support for Programming Languages and Operating Systems}, pp.\
  751--766.

\bibitem[\protect\citeauthoryear{Ludewig and Grieser}{Ludewig and
  Grieser}{2021}]{BMVI2021}
Ludewig, J. and G.~Grieser (2021).
\newblock {Entwurf eines Gesetzes zur Änderung des Straßenverkehrsgesetzes
  und des Pflichtversicherungsgesetzes - Gesetz zum autonomen Fahren}.
\newblock Technical report, No 19/27439, {Bundesministerium für Verkehr und
  digitale Infrastruktur (German Federal Ministry for Digital and Transport)}.

\bibitem[\protect\citeauthoryear{{On-Road Automated Driving (ORAD)
  Committee}}{{On-Road Automated Driving (ORAD)
  Committee}}{2021}]{J3016_202104}
{On-Road Automated Driving (ORAD) Committee} (2021).
\newblock {\em Taxonomy and Definitions for Terms Related to Driving Automation
  Systems for On-Road Motor Vehicles}.

\bibitem[\protect\citeauthoryear{Reke, Peter, Schulte-Tigges, Schiffer,
  Ferrein, Walter, and Matheis}{Reke et~al.}{2020}]{reke2020}
Reke, M., D.~Peter, J.~Schulte-Tigges, S.~Schiffer, A.~Ferrein, T.~Walter, and
  D.~Matheis (2020).
\newblock A self-driving car architecture in ros2.
\newblock In {\em 2020 International SAUPEC/RobMech/PRASA Conference}, pp.\
  1--6. IEEE.

\bibitem[\protect\citeauthoryear{Sari}{Sari}{2020}]{sari2020fail}
Sari, B. (2020).
\newblock {\em Fail-operational Safety Architecture for ADAS/AD Systems and a
  Model-driven Approach for Dependent Failure Analysis}.
\newblock Springer Nature.

\bibitem[\protect\citeauthoryear{Schmid, Schraufstetter, Wagner, and
  Hellhake}{Schmid et~al.}{2019}]{schmid2019safety}
Schmid, T., S.~Schraufstetter, S.~Wagner, and D.~Hellhake (2019).
\newblock A safety argumentation for fail-operational automotive systems in
  compliance with iso 26262.
\newblock In {\em 2019 4th International Conference on System Reliability and
  Safety (ICSRS)}, pp.\  484--493. IEEE.

\bibitem[\protect\citeauthoryear{Thrun, Montemerlo, Dahlkamp, Stavens, Aron,
  Diebel, Fong, Gale, Halpenny, Hoffmann, et~al.}{Thrun
  et~al.}{2006}]{thrun2006stanley}
Thrun, S., M.~Montemerlo, H.~Dahlkamp, D.~Stavens, A.~Aron, J.~Diebel, P.~Fong,
  J.~Gale, M.~Halpenny, G.~Hoffmann, et~al. (2006).
\newblock {Stanley: The robot that won the DARPA Grand Challenge}.
\newblock {\em Journal of field Robotics\/}~{\em 23\/}(9), 661--692.

\bibitem[\protect\citeauthoryear{Urmson, Anhalt, Bagnell, Baker, Bittner,
  Clark, Dolan, Duggins, Galatali, Geyer, et~al.}{Urmson
  et~al.}{2008}]{urmson2008autonomous}
Urmson, C., J.~Anhalt, D.~Bagnell, C.~Baker, R.~Bittner, M.~Clark, J.~Dolan,
  D.~Duggins, T.~Galatali, C.~Geyer, et~al. (2008).
\newblock Autonomous driving in urban environments: {Boss} and the urban
  challenge.
\newblock {\em Journal of Field Robotics\/}~{\em 25\/}(8), 425--466.

\bibitem[\protect\citeauthoryear{Williams}{Williams}{1988}]{williams1988}
Williams, M. (1988).
\newblock {PROMETHEUS-The European research programme for optimising the road
  transport system in Europe}.
\newblock In {\em IEE Colloquium on Driver Information}, pp.\  1--1. IET.

\end{thebibliography}
\end{document}